# IMPACT: INTEGRATED BOTTOM-UP GREENHOUSE GAS EMISSION PATHWAYS FOR CITIES

Juliana Felkner[1,*], Zoltan Nagy[2,*†], Ariane L. Beck[3], D. Cale Reeves[3,5], Steven Richter[2,8], Vivek Shastry[3], Eli Ramthun[2,6], Edward Mbata[1,7], Stephen Zigmund[1], Benjamin Marshall[2], Linnea Marks[4], Vianey Rueda[3], Jasmine Triplett[3], Sarah Domedead[3], Jose Vazquez-Canteli[2], and Varun Rai[3]

[1] School of Architecture, The University of Texas at Austin, TX, USA
[2] Department of Civil, Architectural and Environmental Engineering, The University of Texas at Austin, TX, USA
[3] LBJ School of Public Affairs, The University of Texas at Austin, TX, USA
[4] School of Information, The University of Texas at Austin, TX, USA
[5] School of Public Policy, Georgia Institute of Technology, GA, USA
[6] DNV, Norway
[7] Epic Systems Corporation, Verona, WI, USA
[8] Community and Regional Planning, East Carolina University, Greenville, NC

[*] Equally contributing, co-first authors
† corresponding author, nagy@utexas.edu

## ABSTRACT

Increasing urbanization puts pressure on cities to prioritize sustainable growth and avoid carbon lock-in. Available modeling frameworks fall acutely of guiding such pivotal decision-making at the local level. Financial incentives, behavioral interventions, and mandates drive sustainable technology adoption, while land-use zoning plays a critical role in carbon emissions from the built environment. Researchers typically evaluate impacts of policies top down, on a national scale, or else post-hoc on developments vis-à-vis different polices in the past. Such analyses cannot forecast emission pathways for specific cities, and hence cannot serve as input to local policymakers. Here, we present IMPACT pathways, from a bottom-up model with residence level granularity, that integrate technology adoption policies with zoning policies, climate change, and grid decarbonization scenarios. With the city at the heart of our analysis, we identify an emission premium for sprawling and show that adverse policy combinations exist that can exhibit rebounding emissions over time.

# 1. INTRODUCTION

Buildings account for ~40% of the global energy consumption and ~30% of the associated greenhouse gas emissions, while also offering a 50–90% $CO_2$ mitigation potential (Creutzig et al., 2021; Lucon & Ürge-Vorsatz, 2014; Wang et al., 2018a). Growing urbanization puts pressure on cities in order to absorb increasing populations, requiring decisions on land-use (IEA, 2021; Kennedy, Ibrahim, & Hoornweg, 2014). While it is generally acknowledged that urban infill development is beneficial compared to outward expansion in terms of economics and carbon emissions (Asfour & Alshawaf, 2015; Conticelli, Proli, & Tondelli, 2017; Lima, Scalco, & Lamberts, 2019; McConnell & Wiley, 2012), quantifying these benefits under various external factors, e.g., climate change, is challenging (Teller, 2021). Integration of urban strategies for mitigation and adaptation to climate change is needed to avoid carbon lock-in effects (Seto et al., 2016; Ürge-Vorsatz et al., 2018) and to identify potential synergies and reduce suboptimal trade-offs between mitigation responses. As such, if policies put in place to drive these improvements are to be effective, they should be designed by anticipating the integrated landscape of infrastructure, climate, and behavioral conditions and responses.

End-use electrification in combination with electric grid decarbonization and higher energy efficiency is considered to be the major pathway toward decarbonization of the built environment(Leibowicz et al., 2018). Since energy demand in buildings is mostly dominated by HVAC equipment, one promising policy lever is the provision of financial incentives for higher-efficiency system upgrades or solar photovoltaic (PV) installations (Khanna et al., 2021). Mandates are used in building codes to require that certain minimum efficiency standards or technologies (e.g., solar PV) are met in buildings at the time of building or after major renovations.

It is debated whether individual (bottom-up) action or system-level (top-down) action is more important and should receive greater focus in decarbonization efforts (Goldstein, Gounaridis, & Newell, 2020; Hultman et al., 2020; Khanna et al., 2021). According to the United Nations Environment Programme, however, this is a false dichotomy as both perspectives must be used in conjunction to effect necessary change(United Nations Environment Programme, 2020). The challenges with developing a federal-level coordinated climate policy in the US has no doubt increased focus on state, local, and individual action, but it is unclear how those lower-level actions aggregate to measurable differences in energy use in the urban built environment. Modeling and integrating individual decision-making within the context of changing land-use has become critical to understanding what outcomes we can expect based on undirected individual choice versus those that will require incentives or even mandates to generate the aggregated benefits needed for rapid decarbonization.

Often policies and their impact are evaluated top down, on a national scale, or post-hoc on developments vis-à-vis different policies in the past (Berrill & Hertwich, 2021; Creutzig et al., 2016; Goldstein et al., 2020; Kennedy et al., 2014). Global-scale emission pathway studies typically focus on target warming temperatures and backcast how they can be achieved(Rogelj et al., 2016). Forward projection of emissions and mitigation efforts have only been explored recently in a few studies, considering carbon pricing as a policy mechanism for relatively short target years, e.g., 2030 (Sognnaes et al., 2021) or integration with economic models (Lu et al., 2021). Exploratory forward projections are useful because they can provide realistic estimates of emission reductions under certain given conditions. As a consequence, forward projection can inform what range of emission reductions can be achieved realistically, which in turn could be linked to target warming temperatures.

Here we emphasize that despite cities' position at the forefront for implementation of climate impact mitigation strategies(IEA, 2021; Intergovernmental Panel on Climate Change, 2022), there are no tools available for them to project expected emissions for given policies into the future.

Perhaps the closest related work to our research is the LA100 study (Cochran and Denholm, 2021). It is an example of grid capacity expansion planning with the goal to achieve 100% renewable power grid in the city of Los Angeles, CA, by 2050. The study creates scenarios with assumptions on technology adoption, efficiency and demand flexibility and optimizes for power plant and transmission/distribution costs to achieve 100% renewables. This is a different approach from ours as the target is defined. The study also does not capture decision-making at the individual building level and other demographic factors, and technology adoption is modeled exogenously. In a series of studies (Langevin et al, 2019) and (Langevin et al, 2021) explore the potential of the US building stock to reduce emissions by 80%. The studies explore similar questions to ours, but focus on efficiency measures and power plant planning to reduce emissions. They provide results at the national scale which are not suitable for local planning in municipalities. In fact, their results are aggregated at the US independent system operator (ISO) scale.

These and similar existing models often overlook the integration of key elements like socio-economic factors, zoning policies and consumer-driven technology adoption. Our objective is to bridge this gap with a simulation model that is both comprehensive and user-friendly, specifically designed for policy decision-making. We aim to offer city-level projections, balancing complexity with practical applicability for policymakers. It represents a step forward in providing actionable insights for urban policy formulation and climate strategy implementation.

To address the identified gap in urban planning tools, we present IMPACT: Integrated bottoM-up greenhouse gas emission PAthways for CiTies (Fig 1). IMPACT is designed to analyze the evolution and composition of neighborhoods at a granular level, starting from individual parcels. Each parcel, potentially hosting multiple buildings and residence units, is modeled considering its redevelopment schedule, governed by zoning policies, and its energy demand, dictated by the type of building, the decade, and climate change scenarios. Key to IMPACT is the inclusion of individual 'decision-makers' within each residence unit, who choose whether or not to adopt specific technologies. These decisions are influenced by the available incentives and information, which are themselves outcomes of broader policy and economic scenarios. For instance, in a supportive scenario, financial incentives and mandates drive the adoption of high-efficiency technologies, impacting the energy demand and, consequently, the CO2 emissions of a building. In buildings whose decision-makers decide to adopt high-efficiency/green technologies (in our case HVAC, smart thermostats, solar photovoltaics and storage) the annual energy demand is reduced accordingly. Finally, as a simplification, the energy demand is assumed to be met using fully electrified buildings. This simplification allows us to estimate the resulting CO2 equivalent emissions based on grid carbon content.

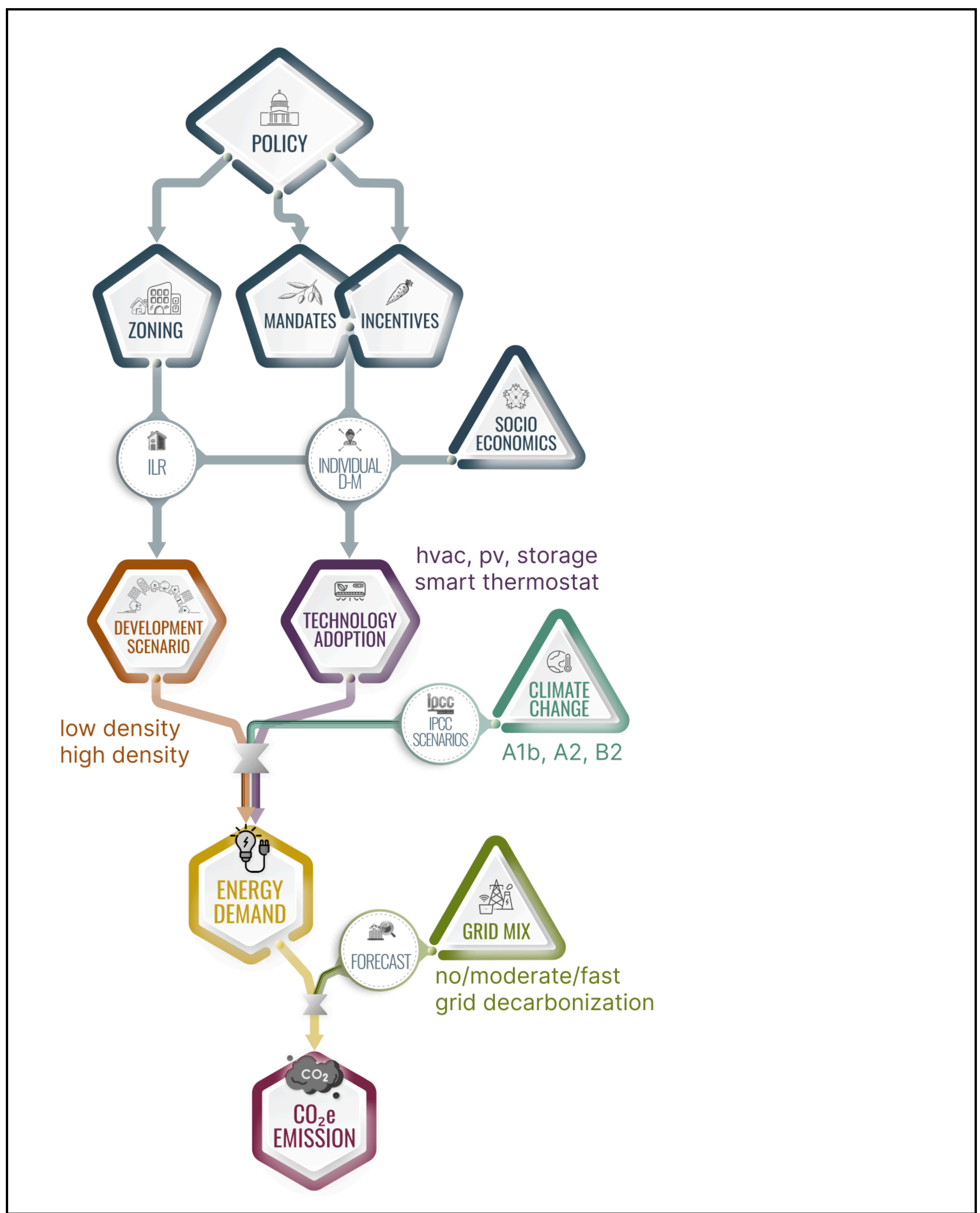

**Figure 1:** Overview of the IMPACT model used to generate multi-domain emission pathways (ILR: Improvement-to-land ratio; Individual D-M: Decision Maker).

Our objective is to create an exploratory model, which for each decade can provide the composition of a neighborhood, e.g., the number of residential, commercial and mixed-use buildings , and its annual energy demand and resulting CO2 emissions under climate change, grid decarbonization and technology adoption scenarios. . Comparing different scenario combinations allows us then to explore how different assumptions play out over longer periods of time, how policies interact and what combinations of strategies provide pathways to reduced operational emissions of neighborhoods. Other metrics and assumptions, e.g., on embodied carbon for construction or transportation emissions, can be easily integrated at the individual parcel and residence level, which makes this a very versatile tool to explore potential pathways and mitigation strategies and policies.

## 2. METHODS

Here, we review the inner workings and assumptions of each component of our model.

### 2.1. Future Land-use and Transformation Scenarios

We apply scenario planning to explore the range of potential energetic impacts associated with changing urban morphology (Schüler, Cajot, Peter, Page, & Maréchal, 2018). We use Envision Tomorrow (ET), an open-access scenario modeling tool, to generate the scenarios(Gabbe & Fregonese, 2013). ET utilizes a set of linked MS Excel spreadsheets with an ArcGIS extension to enable parcel-level land development to be mapped over existing neighborhood geographies, generating demographic, economic, transportation, and energy outputs. ET allows the user to control a range of building and urban design variables, yielding highly differentiated development types necessary for neighborhood scenario development.

We use a parcel-level dataset of the City of Austin, Texas including address, current zoning and land-use class, year of structure construction, assessed property value, as well as an improvement-to-land (ILR) ratio. The ILR is formally defined as the appraised Improvement value divided by the (L)and value and is provided by the city assessors, i.e,. ILR=I/L where the Improvement value is the assessed value of the structure. The ILR can be interpreted as a measure of the economic potential of a property (City of Austin, 2009). We use the ILR to create a parcel rank of re-development likelihood used to schedule parcel redevelopment through 2100, where we assume that a parcel is more likely to redevelop the smaller its ILR is.

We selected study neighborhoods within Austin, Texas, USA, guided by the likelihood of a neighborhood experiencing major changes to building morphology due to (re)development. Given Austin's high rates of population growth, housing demand, and resulting increasing residential property values, neighborhoods currently composed largely of older single-family homes were seen as the most likely candidates to experience major redevelopment. Further narrowing criteria for identifying neighborhoods included homes constructed before 1970 (as an indicator of opportunity for upgrades or replacement), the relationship between lot size and existing building footprint (as an indicator of under-utilization), and the existence of households below the median income of Austin (as an indicator of gentrification pressure). Finally, we included geographic variation as a factor to account for property value differentials that might impact redevelopment. The three selected neighborhoods are Brentwood, South Menchaca, and Montopolis, and provide a diverse geographic, income, construction age, infill and redevelopment potential based on the above criteria. Specifically, Brentwood, South Menchaca, and Montopolis represent a progression from most-to-least utilized and, conversely, least-to-most vulnerable to gentrification. **Table 1** shows the characteristics of the three neighborhoods at the beginning of our modeling period in 2020.

| | Nr of Lots | Total Lot Area ($m^2$) | Housing Units | Total Housing Area ($m^2$) | Average 2020 population |
|---|---|---|---|---|---|
| **Brentwood** | 2,580 | 3,340,000 | 4,790 | 606,240 | 14,350 |
| **South Menchaca** | 2,378 | 2,831,000 | 3,099 | 441,520 | 9,100 |
| **Montopolis** | 2,143 | 4,744,000 | 3,619 | 494,530 | 11,600 |

**Table 1** Characteristics of the studied neighborhoods

We envision two development pathways: a) a low-density, sprawling, future with redevelopment indicative of auto dependent urban patterns, consisting primarily of larger single-family homes, and higher floor area per capita(Pincetl et al., 2014) and b) a high-density future supported by greater pedestrian activity and transit (Appleyard, Ferrell, & Taecker, 2017; Cervero & Landis, 1997) with high intensity, multi-story residential and commercial buildings along major streets, and greater reliance on intermediate density multi-family ("missing middle"). The future land-use scenarios for each of the three neighborhoods were developed starting in 2020 for the target year 2100 in steps of 10 years with the following assumptions: First, we assume that all buildings will be redeveloped by 2100 and, second, that each parcel will only go through a single redevelopment. These assumptions serve to simplify the scenario process for the purposes of assessing the impacts of morphology on future energy use, in our exploratory scenarios. We also create a reference scenario that reflects the current land-use, i.e., no redevelopment.

To account for the uneven nature of land redevelopment pressure, the future land-use scenarios contain internal differentiations by land-use class and location. Specifically, we initially classified the parcels based on their current land-use and location. We combine the current land-use into three classes, representing similar building morphologies: small residential, large residential, commercial/mixed-use. Similarly, we identify three categories to account for location: along a major traffic corridor, within an identified transit-oriented development (TOD) area, or within the interior of the neighborhood. Major corridors and TOD areas were determined by the *Image Austin Comprehensive Plan*, which specially identified the locations as preferred "activity corridors" and "centers" for future growth.(Wallace, Roberts, & Todd, 2012). Class (3), location (3), and scenario (2) combine to form 18 lookup tables (3 x 3 x 2) where parcel size differentiates which building within the lookup is applied to each redeveloped parcel. The general strategy is for larger parcels to have more dwelling units and square footage. Flag fields in the parcel database identify which lookup table a given parcel reads from.

We build a redevelopment schedule into the dataset to explore the process of land-use changes by decade. The redevelopment schedule is the percentage of parcels in the neighborhood considered to redevelop over the course of each decade from 2020 to the target year of 2100 (see **Table 2**). Montopolis has large, undeveloped portions, and therefore a great percentage of parcel redevelopment occurs quickly. The redevelopment schedule is implemented through a redevelopment rank determined by the ILR and broken out by the three classes (small res, large res, mixed use ). The lowest ILR in each location is assigned the highest rank for each use class. Redevelopment is implemented using the percentage of all parcel land-use ordered by max rank by decade. The major differences in the development of the different neighborhoods can be summarized as

- Montopolis (Mont) - the only neighborhood with significant greenfield potential. We chose to subdivide the larger lots inside the neighborhood since these are likely to become small residential in the near term (no other subdividing or aggregation occurred). But there remain many large lots suitable for mid-rise development. Existing single-family lots are the biggest of the three. So: greatest potential variability neighborhood.
- Brentwood (Brent) - the smallest lot sizes, both for larger and smaller parcels. But has significant acreage of larger parcels suitable for mid-rise development (especially compared to SM). So: scenario leaning toward concentrated density
- South Menchaca (SM) - larger lots than Brent, but a greater share of small residential parcels. The least potential for mid-rise development. So: more decentralized density (8-plexes inside the neighborhood)

| | 2020 | 2030 | 2040 | 2050 | 2060 | 2070 | 2080 | 2090 | 2100 |
|---|---|---|---|---|---|---|---|---|---|
| **Montopolis** | 15% | 15% | 6% | 6% | 9% | 9% | 10% | 15% | 15% |
| **Brentwood** | 6% | 6% | 6% | 9% | 9% | 12% | 12% | 20% | 20% |
| **South Menchaca** | 6% | 6% | 6% | 9% | 9% | 12% | 12% | 20% | 20% |

**Table 2:** Redevelopment schedule (% of lots redeveloped) in each neighborhood

### 2.2. Architectural and Energetic Modeling of Building Archetypes

Envision Tomorrow (ET) has precomputed constant annual energy demand for each of its various building archetypes, which is not suitable for our purposes because it does not reflect impact of climate change or efficiency upgrades. Thus, we develop the equivalent building energy models for the 25 different residential and commercial building archetypes from ET (Figure 2 ), which are then subsequently used in the land-use and transformation scenarios described above. We have made several assumptions to make the models realistic and consistent with the ET archetypes. For example, d etached homes are assumed to have rectangular floor plans 7.6m width and the length adjusted to the area assumed in ET. Ceiling heights are 3m. For multi-family and multi-use building types, we assume double loaded 2.4m corridor down the middle, and the different units are studio, one-, two-, three- and four-bedroom apartments with a standard depth of 7.6m and the length again adjusted according to the overall area. Mixed-use buildings have storefronts at the bottom floor. All buildings assume a wood frame construction, and double pane windows. We assume 2.5 people per residential dwelling unit. The buildings are designed in Rhinoceros 3D(McNeel, 2010) and simulated with EnergyPlus(Crawley et al., 2001) using the DIVA plugin (Jakubiec & Reinhart, 2011) and the weather files corresponding to the three studied climate change pathways.

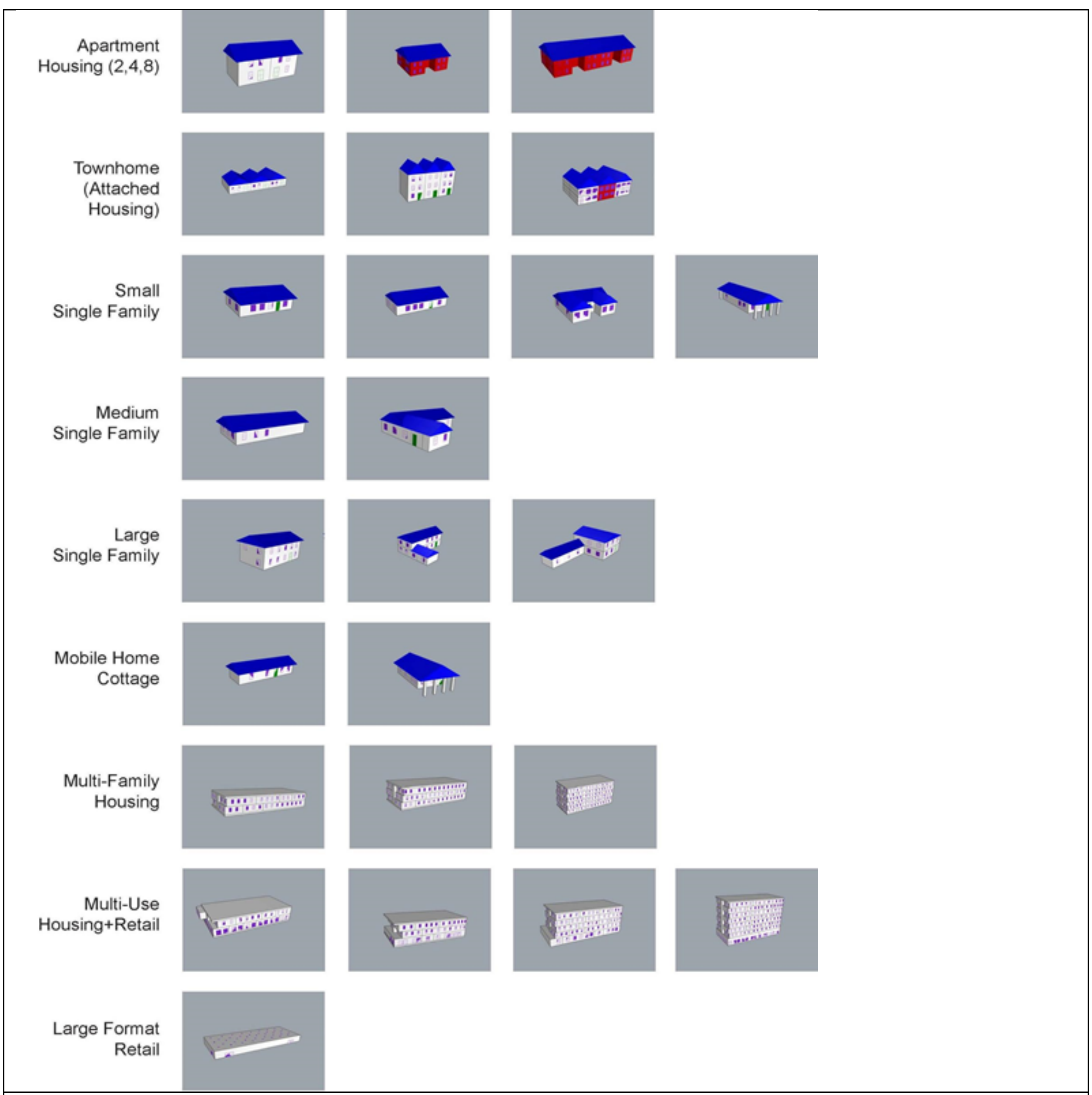

**Figure 2 Building archetypes**. Detached homes are assumed to have rectangular floor plans 7.6m width and the length adjusted to the area assumed in Envision Tomorrow. Ceiling heights are 3m. For multi-family and multi-use building types, we assume double loaded 2.4m corridor down the middle, and the different units are studio, one-, two-, three- and four-bedroom apartments with a standard depth of 7.6m and the length again adjusted according to the overall area. Mixed-use buildings have storefronts at the bottom floor. All buildings assume a wood frame construction, and double pane windows. We assume 2.5 people per residential dwelling unit.

## 2.3. Climate Change Pathways

For the energy simulation under climate change, special weather files are required that have projections for a typical weather year in each decade for a given climate change scenario. W e use weather files for Austin, TX based on the IPCC 2000 Special Report on Emission Scenarios (SRES) provided by Meteonorm (Meteonorm, 2019) , in addition to the typical mean year (TMY) weather data for current weather reference. The SRES scenarios make projections on future warming based on a combination of economic and environmental parameters as well as pathways towards more or less globalization. Meteonorm provided us with the three scenarios A1B (rapid economic growth in a globalized world), A2 (regionally oriented economic development) and B1 (global environmental sustainability). These SRES scenarios are similar for our purposes (assess the impact of warming) to the more recently published Representative Concentration Pathways (RCP) and shared socioeconomic pathways (SSPs) scenarios. For example, SRES A1B is similar to RCP 6.0, SRES B1 is similar to RCP 4.5, while SRES A2 is between RCP 6.0 and 8.5 (Riahi et al., 2017). Given that we are only interested in the warming, the three scenarios collectively show a breadth of potential outcomes. The three SRES climate scenarios (B1, A1B and A2) result in an average annual temperature increase in 2100 in Austin of +1.5C, +2.5C and +3C, respectively.

### 2.4. Policy Instruments for Technology Adoption

Currently in Austin, TX the following policy incentives are available for the technologies being considered in this model. For Solar PV, homeowners can avail a Federal Investment Tax Credit (FITC) of 26%, which was stepped down from 30% in 2020. Since 2019, a flat rebate of $2,500 is also available for systems over 2.5kW. For upgrading to efficient HVAC, a rebate of up to $2550 is available since 2020, whereas installing a smart thermostat can earn a $110 rebate. From these existing baselines, we model two bounded policy scenarios for technology adoption. In the best-case scenario: 1) the FITC for solar PV steps down to 22 percent in 2023 and expires in 2024 as defined by the current federal policies; and 2) the rebates for all the three technologies continue to exist until 2100, with the rebate for solar PV available for systems above 1.2kW from 2022 onwards. In the worst-case scenario, the FITC, as well as all the local rebates, expire by 2020 and no economic incentives are available for any of the three technologies under consideration. Modeling these two scenarios provides upper and lower bounds for adoptions, which allows the integrated model to explore the full range of impact from individual adoption decisions.

### 2.5. Agent-based model of technology adoption in households

We model the diffusion of energy technologies at the household level using an agent-based model (ABM) as demonstrated previously (Rai & Robinson, 2013; Robinson & Rai, 2015). The ABM approach allows us to simulate realistic social drivers of home energy technology adoption decision-making. At the core of the model is a dual-threshold model of *gateway* technology adoption mated to a novel *sequential* model of technology co-adoption. The gateway model identifies the first technology adopted by a household as a function of their access to financial and informational resources and the conditions at the time of their financial and informational activation. The sequential model specifies the order and timing of subsequent co-adoptions conditioned on prior adoptions – including gateway technology – and dwelling type. Note that single family homes can adopt all possible technologies (high efficiency HVAC, solar, storage and smart thermostats), while multi-family homes can only adopt smart thermostats. The available roof space for PV is estimated from building footprint, tree cover and elevation; orientation of the roof is not explicitly estimated.

The ABM initializes with more than 181,000 buildings, including single-family homes in Austin, Texas, as well as single- and multi-family homes within focal neighborhoods, depending on the development scenario. Agent state data at initialization includes indices for access to financial and informational resources, geospatial location, and type of dwelling. Agents' social networks are estimated at initialization in three steps: 1) all alter agents within a geographic radius $\varphi$ of an ego agent are geographic candidates for connection, 2) a homophily constraint – the top $\rho$ percent of similar agents according to financial resource access – is applied to the geographic candidates and the remaining agents are connected, and 3) an additional $\lambda$ proportion of the total connected neighbors are randomly selected from the entire agent pool and connected. The resulting social network is empirically informed and has small-world characteristics.

The economic and policy context are also established at initialization. The economic context consists of future sale prices of home energy technologies estimated by combining historical data with simple trend assumptions, e.g., stable decreases in prices over time with constant variability. The policy context comprises the primary set of decision variables in the model and includes many aspects that shape agent decision-making. For example, the rebate available for any home energy technology at any point in time reflects a policy interest in offsetting a portion of the financial burden (captured in the economic context) associated with technology acquisition. Similarly, a mandate requiring that all new units have a particular technology – regardless of rebate availability – reflects policy interest in that technology.

The informational context captures the social drivers of technology adoption: during the simulation, agents exchange information with their social neighbors, altering the level and distribution of information in the system. As the simulation progresses, agents make adoption decisions that diffuse the target technologies. Agents are sparked to adopt a gateway technology when they acquire sufficient informational resources: i.e., when they are convinced that adopting the technology is a good idea through the dynamic and emergent informational context. The gateway technology and subsequent technologies that compose the agents' home energy plans are randomly selected from the empirical distribution of gateway technologies and subject to constraints imposed by agent status as renter or owner. Once activated with a gateway technology, agents solicit bids to install each technology in the home energy plan. Successful adoption occurs when agents solicit bids that they can afford: i.e., the agent has access to sufficient financial resources as indicated through their financial index. When weighing the prospective benefits of a bid, the benefit of each technology is calculated with respect to the suite of previously installed technologies. For example, the benefit of installing a smart thermostat differs for agents who do versus do not already have solar PV installed; in the first case, it would reduce their overall energy use, which is valued at the rate of the feed-in-tariff, while in the second, its reduction in energy use would be valued at the retail electric rate.

Also, as the simulation progresses, development scenarios determine agent exit from, and entry to, the population. When a scenario includes changes in a parcel's use (e.g., density changes such as shifting from a lone single-family home to two single-family homes on the same parcel), the ABM creates and removes agents as appropriate. New agent states are initialized following the procedure described above.

### 2.6. Efficiency improvements of adopted technologies

We estimate the effect of energy efficiency measures as follows. If a building is adopting High-efficiency HVAC, it's annual energy demand is reduced by the highest efficiency available for that year based on the technology adoption model; if a building is not adopting high-efficiency technologies, it follows a regular lifetime update, i.e., every 20 years, the HVAC system is updated with one that is slightly more efficient, with the general efficiency improvements assumed to be 2% per year for 20 years until a theoretical limit is reached(Wang et al., 2018a ).

The effect of solar PV and storage on the energy demand is estimated in two steps. First, we determine the annual energy generated $E_a$ in (kWh) based on the panel size that has been selected by the technology adoption module using on the solar sun hours method as

$$E_a = 356 \times S_{avg} \times P$$

Where $S_{avg}$ (in hr) is the average daily sun hours in a location, and $P$ in (kW) is the nominal power output of the solar array. For Austin, $S_{avg} \approx 5$ hr. The impact of the added battery is modeled by assuming an average annual self-sufficiency of 40%. Using these numbers for energy improvements, each building's pre-simulated energy demand is updated and reduced to reflect technology adoption.

## 2.7. Grid Decarbonization Scenarios

We include three grid mix evolution pathways in our study. The first maintains the 2020 level of carbon content at ~430 $gCO_2eq/kWh$ (no grid decarbonization) and serves as a reference. This number is the average for the grid mix from 2010-2019 (City of Austin, 2022). As an indication, the 2019 grid mix was approximately 47% natural gas, 20% coal, 20% wind, 11% nuclear, and 1% wind (ERCOT, 2019) . A rapid grid decarbonization scenario is used for the TX grid (Rhodes & Deetjen, 2021), at a rate of about -100 $gCO_2eq/kWh$/decade reaching a constant value of ~50 $gCO_2eq/kWh$ by 2060. This is a "net-zero by 2050" scenario as the final value of 48 $gCO_2eq/kWh$ by 2100 represents the embodied emissions of the renewable generation in the grid(Intergovernmental Panel on Climate Change, 2014). The third, moderate, pathway is defined as the arithmetic average between the previous two, resulting in a decarbonization rate of –50 $gCO_2eq/kWh$/decade until its plateau at about 240 $gCO_2eq$ in 2060, after which grid decarbonization efforts stall at a non-zero operational emission grid mix. These three pathways cover a large variety of overall carbon content in the TX grid, regardless of the generation composition, which is sufficient for our case. The grid carbon content factor is used to convert annual operational electricity demand to annual greenhouse gas emissions.

## 2.8. Comparing high and low-density development: Premium for Sprawl

We are generally interested in assessing the emissions in the buildings belonging to the study neighborhood, i.e., limited by their geographical area. However, keeping the area fix, means that densification allows more people to be accommodated, who otherwise would move to another area and whose emissions in absolute terms would not be counted. Thus, to allow a fair comparison in absolute emission values between different urban developments, we extrapolate the results for the low-density developments to match the number of units of the high-density development. We then define the *Premium for Sprawl* as the difference in emissions (in $tCO_2eq/yr$) of low-density developments compared to high-density developments. In other words, the Premium for Sprawl describes the surplus in emissions due to sprawling for a fixed number of residences (or persons). The *Premium for Sprawl* is inspired by the *Premium for Height* for tall buildings, resulting from the increase of cost for the required material to withstand wind loading (Ali & Moon, 2007).

### 2.9. **Model Limitations**

To convert energy demand to emissions in the buildings, we assume that all energy used is electric, i.e., all buildings are electrified. For our case study in Texas, where cooling is the dominant energy use and typically met with electric air conditioning systems, this is a reasonable assumption, especially to compare pathways amongst each other. For other climates, where heating is dominant, one must include fuel switching scenarios that also consider transitioning from fossil fuel heating systems to electric heating systems, e.g., heat pumps.

Urban energy systems, such as district heating and cooling are not investigated. Also, our models are not capturing extreme weather situations like heat waves or cold snaps. As indicated above, our model also does not include emissions from transportations or embodied construction emissions. However, the parcel, level formulation of the model allows the integration of these with reasonable assumptions in the future. Similarly, if a realistic initial condition of the neighborhood in terms of construction material can be created, then building retrofitting, e.g. envelope and window improvement would be another scenario dimension that could be explored (Felkner & Brown, 2020).

As with any long-term forecasting models, we assume that general behaviors, e.g., on technology adoption or urban transformation drivers do not change significantly over time. While these are strong assumptions, they also let us investigate their relative importance, such that one can decide which of the models should be further improved to better assess their overall impact.

We are not explicitly including population numbers in the neighborhoods. Instead, we couple population numbers to the building units by assuming 2.5 occupants per residence, which is consistent with the US average household size according to the 2018 census. As a consequence, per unit indicators are equivalent, in our case, to per capita indicators.

## 3. RESULTS: IMPACT PATHWAYS

### 3.1. Importance of Baseline

One important aspect of policy development and interpretation is the choice of the appropriate baseline against which potential decarbonization interventions are compared. Recently, it has been argued that policies to address climate change should be analyzed by considering only the outcome when climate change is accounted for (Hausfather & Peters, 2020; Jafino, Hallegatte, & Rozenberg, 2021). This helps capture potential interactions and avoids overestimating the impacts of the policy. We illustrate this in Figure 3, where we show the annual energy demand of the neighborhoods for the climate change forecasts compared to their current state (2010) and excluding urban redevelopment and technology adoption (grid carbon content is irrelevant for energy demand).

Energy demand is increasing by +10% by 2050 in any climate scenario and increases to above 20% by 2100 for A1B and A2, while plateauing at about +15% for B1 after 2080. Therefore, any energy efficiency policy that fails to include climate change, will overestimate its expected impact by at least 10% by 2050. IMPACT pathways reduce this bias by incorporating climate change scenarios. While this may seem rather obvious, surprisingly few, if any, building energy and decarbonization scenarios are presented with climate changed modified weather scenarios.

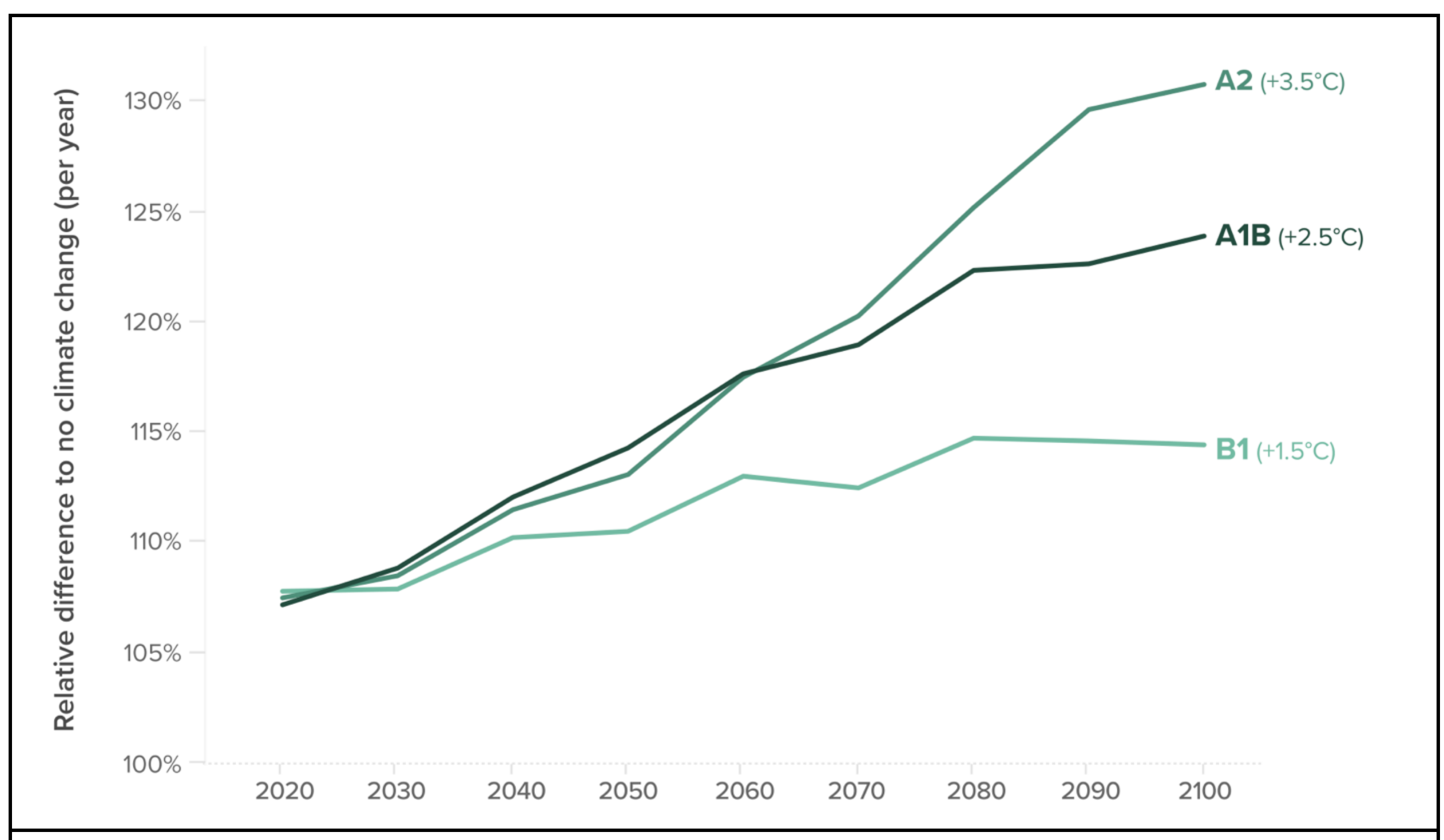

Figure 3 Energy demand under climate change. It is crucial to integrate changing climate weather data into future energy projections. Energy efficiency policy that fails to include climate change, will overestimate its expected impact.

### 3.2 . IMPACT: Policy Interactions: Synergies, Trade-offs and Rebounds

We now explore the IMPACT pathways for the three neighborhoods in Austin, TX with the scenarios shown in Figures 4 and 5. Each individual pathway is a combination of the urban development scenarios (no change, high density, low density), grid decarbonization (no, moderate or rapid decarbonization), and technology adoption (no adoption, neutral, and supportive policy). The combinations are labeled with uppercase letters (A-J) in Figures 6/7 and shown in the Figure legend. For clarity, we only show climate scenario A1B and aggregate the emissions for all neighborhoods. We will discuss the implication of climate change itself further below. Scenarios A/B consider rapid decarbonization for low and high density developments, respectively. Similarly, scenarios D/E consider moderate grid decarbonization, and C/G assume no grid decarbonization. Notice that the same letter refers to the same scenario in both figures to allow for comparison. Finally, scenarios F, I and J are reference scenarios that only consider climate change and moderate (I) and rapid (J) grid decarbonization, i.e., no housing development and no technology adoption, while scenario F only considers climate change.

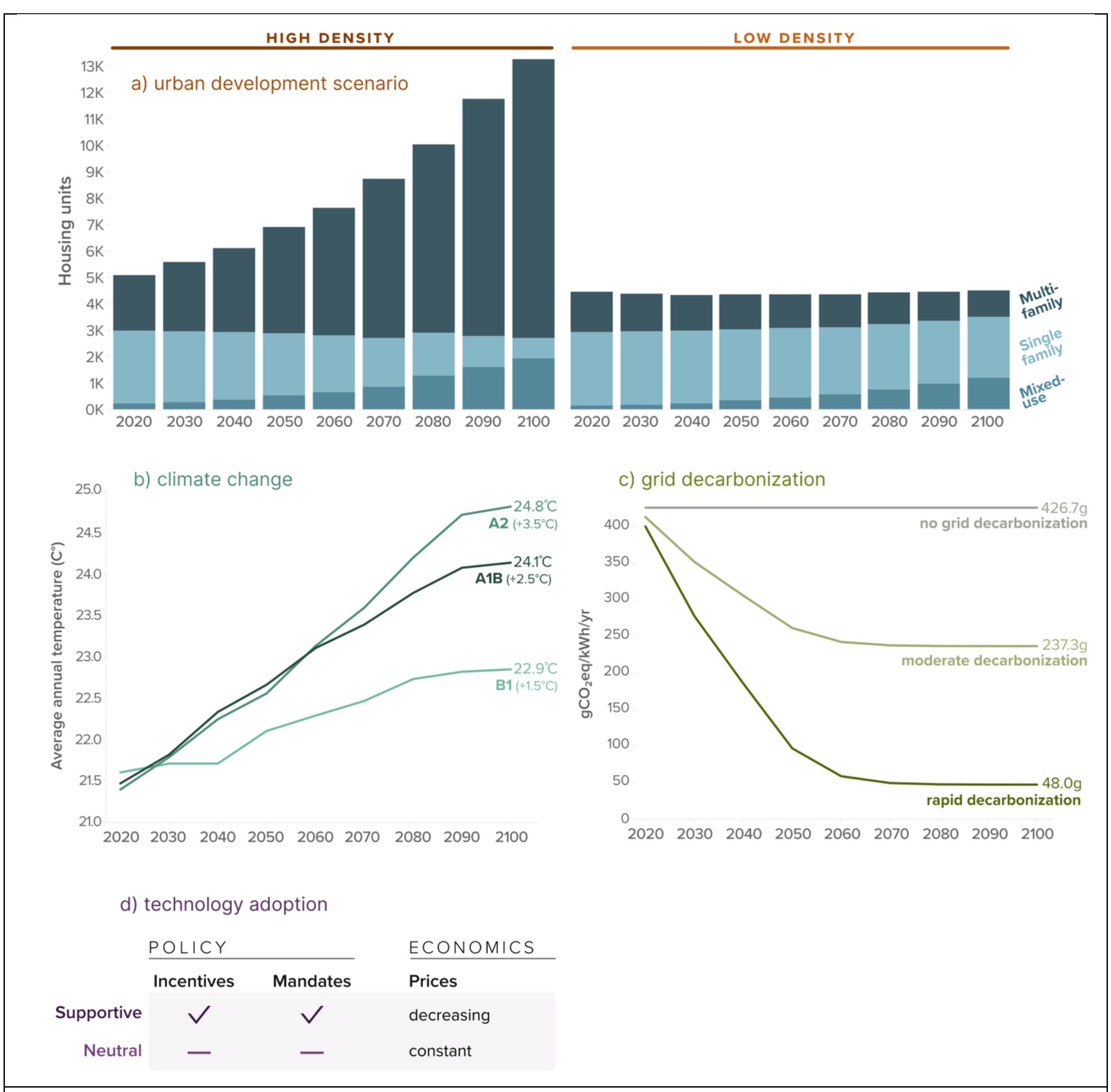

Figure 4 Scenarios for our case study: a) Number of building types for high- and low-density urban development scenarios for the Brentwood neighborhood in Austin, TX b) Average annual temperature in Austin, TX, for different climate change scenarios, c) grid decarbonization scenarios, and d) Policy and economic inputs for technology adoption scenarios

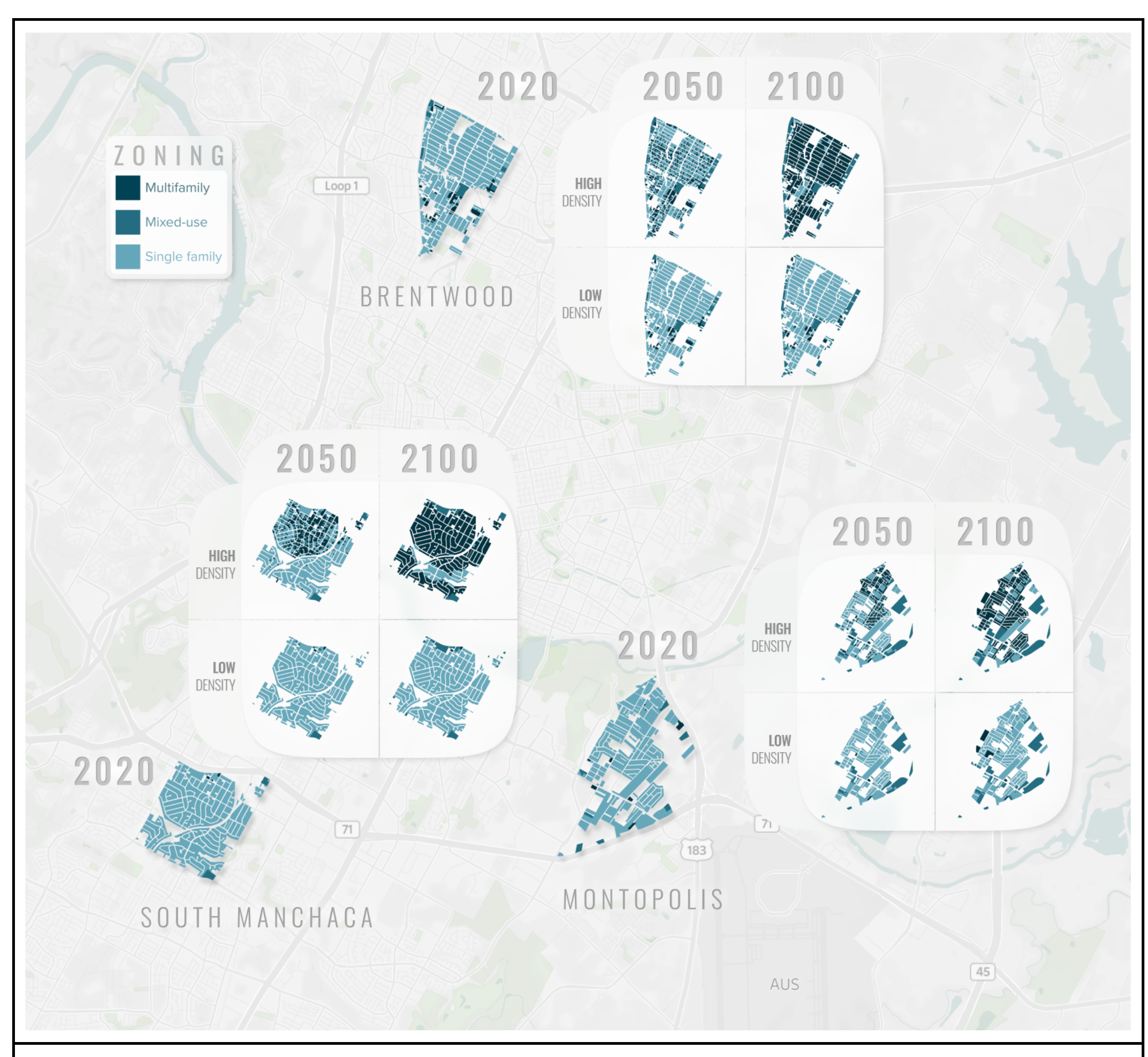

**Figure 5: Residential zoning policy and urban development scenarios in each of the three studied neighborhoods**

Figure 6 and Figure 7 and show the relative (per residence unit) and absolute IMPACT pathways, respectively, for the A1B climate scenario and all other considered scenarios, aggregated for all three studied neighborhoods (shown in Figure 5 ).

The relative pathways offer an apples-to-apples comparison between the urban development scenarios (**Figure 6** ). Clearly, fast grid decarbonization has the largest overall effect on emission reductions (**Figure 6** (A/B/J)). In addition, both low- and high-density development further amplify the emissions reductions in the beginning (**Figure 6** (A/B)). However, after 2070 the emissions of the low-density development begin to slightly increase again (Fig.6 A), while they remain flat for the high-density development (Fig. 6 B).

In the moderate grid decarbonization scenario, the low-density development *rebounds* its emissions after 2050 (Figure 6 (D )), while for densification the annual emissions remain flat after 2050 (**Figure 6** (E )). In fact, densification without grid decarbonization (**Figure 6** (C )), reduces the relative emissions of the neighborhoods by about 25% between 2020 and 2100. By contrast, low-density development without grid decarbonization first reduces the annual emissions by 20% in 2050, but ultimately *rebounds* by 2100 to about the same levels as 2020 (**Figure 6** (G )). While this is about 20% lower than the reference case that only considers climate change and no urban development or grid decarbonization (**Figure 6** (F )), it is also 30% higher than the corresponding high-density development (**Figure 6** (C )).

Comparing **Figure 6** (C ) and (I ), we observe that several scenario combinations can lead temporarily to similar outcomes: no grid decarbonization with high density development (**Figure 6** (C )) and moderate grid decarbonization alone (**Figure 6** (I )) have about the same annual emission until about 2050, after which they diverge. Thus, the same relative emission pathways can be achieved with different policy combinations.

Our results clearly demonstrate that in the absence of a zoning policy that is favorable for densification, the major driver for the decarbonization of the neighborhoods is grid decarbonization. This is a rather important realization as the drivers behind the two are not necessarily related or combined and subject to different socio-techno-economic and political boundary conditions. We demonstrate here that their interaction has a substantial effect on emissions outcomes and pathways.

Our results also show that technology adoption has a comparatively small impact, e.g., the three curves for **Figure 6** (D ) representing the neutral, no tech adoption and supportive policies have almost identical pathways. Therefore, all technology adoption scenarios are implicitly assumed in the corresponding grid decarbonization and urban development scenarios.

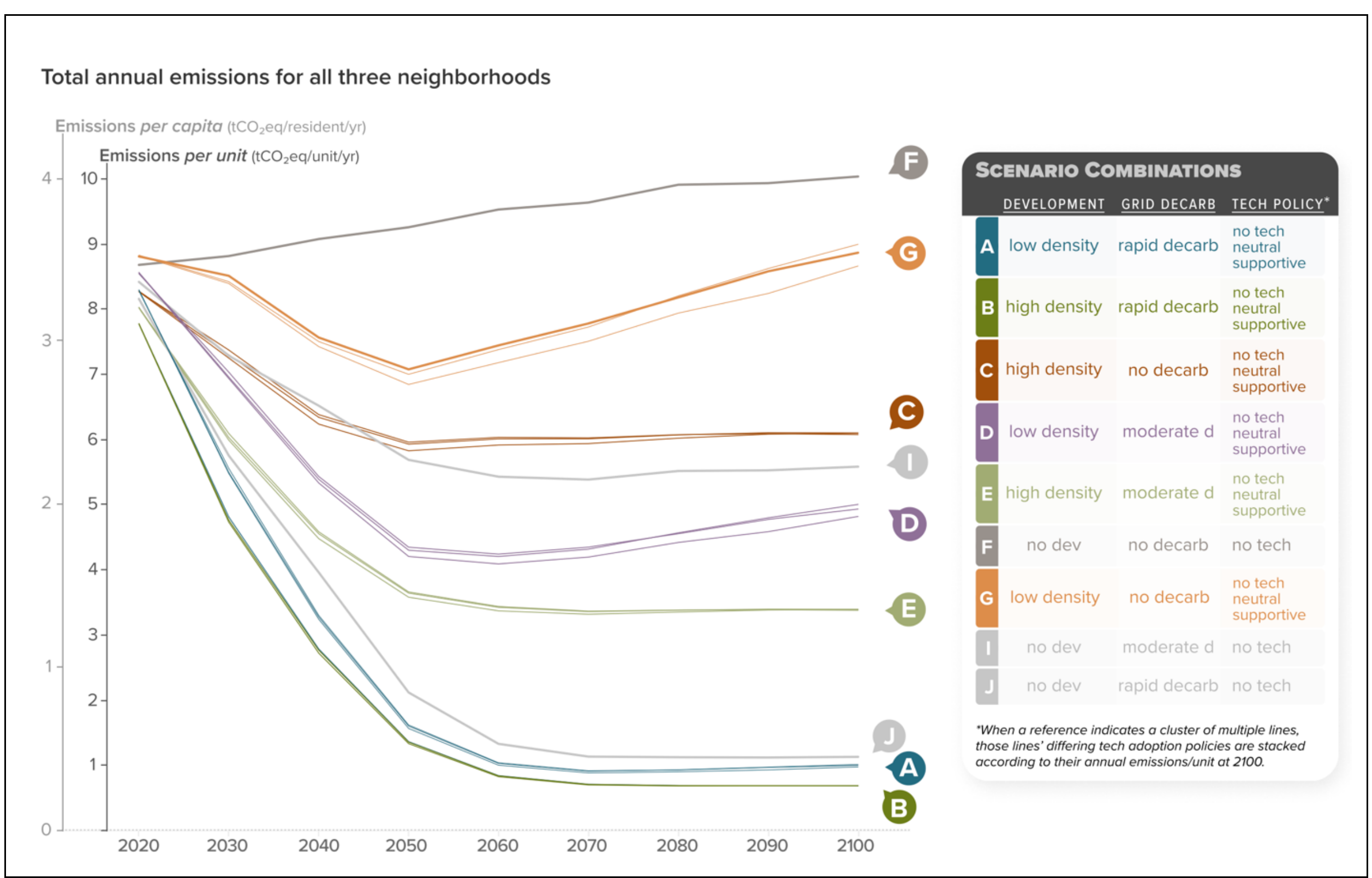

**Figure 6 : IMPACT Pathways for relative annual emissions aggregated for all three neighborhoods and climate scenario A1B.** Rapid grid decarbonization results in the fastest emission reductions (A/B/J) with densification (B) further amplifying emission reductions. For moderate grid-decarbonization, low-density development (D) shows rebound of emissions after 2060, while high-density development (E) does not show rebound. Densification without grid decarbonization (C) can reduce emissions without rebound and is equivalent in reduction potential until 2050 to moderate grid decarbonization alone (I). Low density development without grid-decarbonization (G) reduces emissions until 2050, but rebounds by 2100 to higher emissions than 2020. Finally, without any decarbonization measures, emissions increase due to climate change (F)

In terms of absolute emissions (**Figure 7** ), again rapid grid decarbonization of the grid leads to the fastest emission reductions by far, regardless of other policies (A/B/J). Because there are fewer buildings in the low-density neighborhood, its annual emissions in this case (**Figure 7** (A)) are somewhat lower than the corresponding high-density neighborhood (**Figure 7** (B)). At the other grid decarbonization extreme, i.e., no grid decarbonization, densification and climate change substantially increase the overall emissions (**Figure 7** (C)) due to the increased number of units in the neighborhoods. This somewhat counter intuitive result stems of course from the fact that the high-density neighborhood absorbs many more people, which are not considered in the low-density scenario. We further discuss and compare high and low density developments below in Section 3.4.

For moderate grid decarbonization and low-density urban development (**Figure 7** (D)), grid decarbonization mainly drives the initial emission decrease. After 2050, however, emissions begin to *rebound* and by 2100 the annual emissions return to their level of about 2040. The high-density development even rebounds to annual emissions higher than their 2020 level (**Figure 7** (E)).

If only climate change is considered (no grid decarbonization, and no urban redevelopment), the annual emissions increase slightly (**Figure 7** (F)). Adding low-density urban development (**Figure 7** (G)) shows that it can reduce emissions until about 2050, due to efficiency increases in newly built buildings. However, here also the emissions eventually rebound, due to energy demand increase driven by climate change, and by 2100, the annual emissions return to their values at about 2020.

Comparing **Figure 7** (E) and (G), we again see that different scenario combinations can temporarily achieve the same emission outcomes: Both set of curves (E: moderate grid decarbonization and high-density development) and (G: no grid decarbonization and low-density development) follow a similar reduction until about 2050, and a similar rebound until about 2070. After 2070 their pathways separate. Notice that this is contrary as for the relative pathways, where the two scenario combinations are clearly separated .

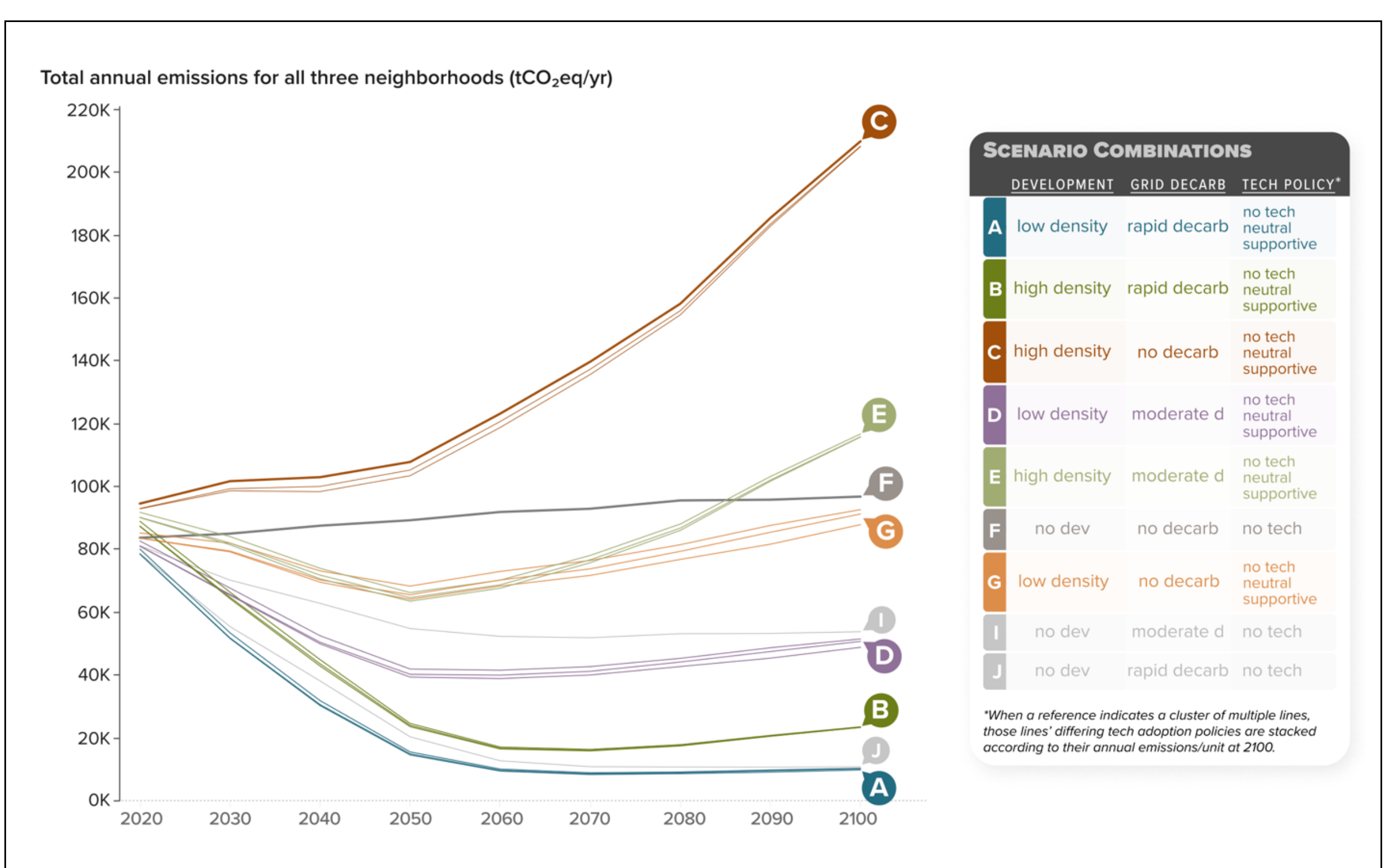

**Figure 7 : IMPACT Pathways for absolute total annual emissions aggregated for all three neighborhoods and climate scenario A1B.**

Rapid grid decarbonization results in the fastest emission reductions (A/B/J). In absolute terms, the high-density developments (B, E and C) show higher emissions compared to their low-density counterparts (A, D and G) in any grid decarbonization scenario.This is due to the increased number of units or population. Most scenarios show some form of emissions rebound after 2050 regardless of densification (E, G, D, B), except for rapid decarbonization (A/J) which is the only scenario that sustains low emissions, and high-density development only (C) which only increases emissions. Scenarios E and G overlap substantially until 2060 and diverge afterwards showcasing the potentially long-term impact of policy measures.

### 3.3 . Technology adoption policy vs Urban Development

Here, we are interested in comparing the impact of technology adoption to urban development policies for a given grid decarbonization. We use the moderate grid decarbonization scenarios for this (Figure 6 D and E). **Figure 6** shows that urban development has a substantially larger effect on emission reductions compared to technology adoption. Clearly, technology adoption reduces emissions in all cases . The effect is larger for the low-density development, where sustained reductions can be achieved during the rebound phase (**Figure 6** (D )). This is because u nder high-density development, single-family dwellings give way to multi-family dwellings – for which the full menu of technology adoption adoptions is typically not available – effectively eliminating the impact of technology adoption over time (**Figure 6** (E )). Of course, because overall accumulated emissions matter more for slowing climate change, every bit helps.

This has clear implication in the ongoing policy discourse of individual/market-driven solution (e.g. technology adoption in our model) vs systemic solutions (here: urban development). Clearly, technology adoption incentivization alone cannot be the central cornerstone of any serious climate policy. We conjecture that the same conclusion could be drawn for other types of technology adoptions. For example, we do not explicitly model fuel-switching adoption, i.e., a home switching from a gas-furnace for heating to an electric heat pump. However, given that that is also typically a high price upgrade, the adoption rate would be similar to what is presented here.

### 3.4 . Climate change and Urban Development

As the emission pathways above (Figs 6 and 7) only consider the A1B climate scenario, here we revisit the impact of all three scenarios (A1B, A2 and B1) and a no climate change baseline, while assuming the moderate grid decarbonization scenario and no technology adoption. Figure 8 a shows the impact of climate change on the relative emissions ($tCO_2eq$/unit) for each urban redevelopment scenario. As the first part (until 2050) is driven by grid decarbonization and the different climate scenarios are still relatively similar, there is virtually no difference within the urban development scenarios , low-density redevelopment having higher relative emissions than high-density, and both being higher than their emissions without climate change. After 2050, as grid decarbonization stalls in the assumed moderate scenario, and climate change intensifies , different pathways emerge

As one would expect, the A2 climate scenario has the highest associated emissions, while B1has the lowest. We can see that an amplifying high emission combination pathway (A2 & low-density development) can be up to 65% higher than an amplifying low emissions combination pathway (B1 & high-density development) in 2100. Further, a favorable climate scenario (B1) coupled with low-density development is still about 30% higher in 2100 than the worst-case climate scenario (A2) with high-density development. In other words, an unfavorable zoning policy will be considerably amplified by climate change.

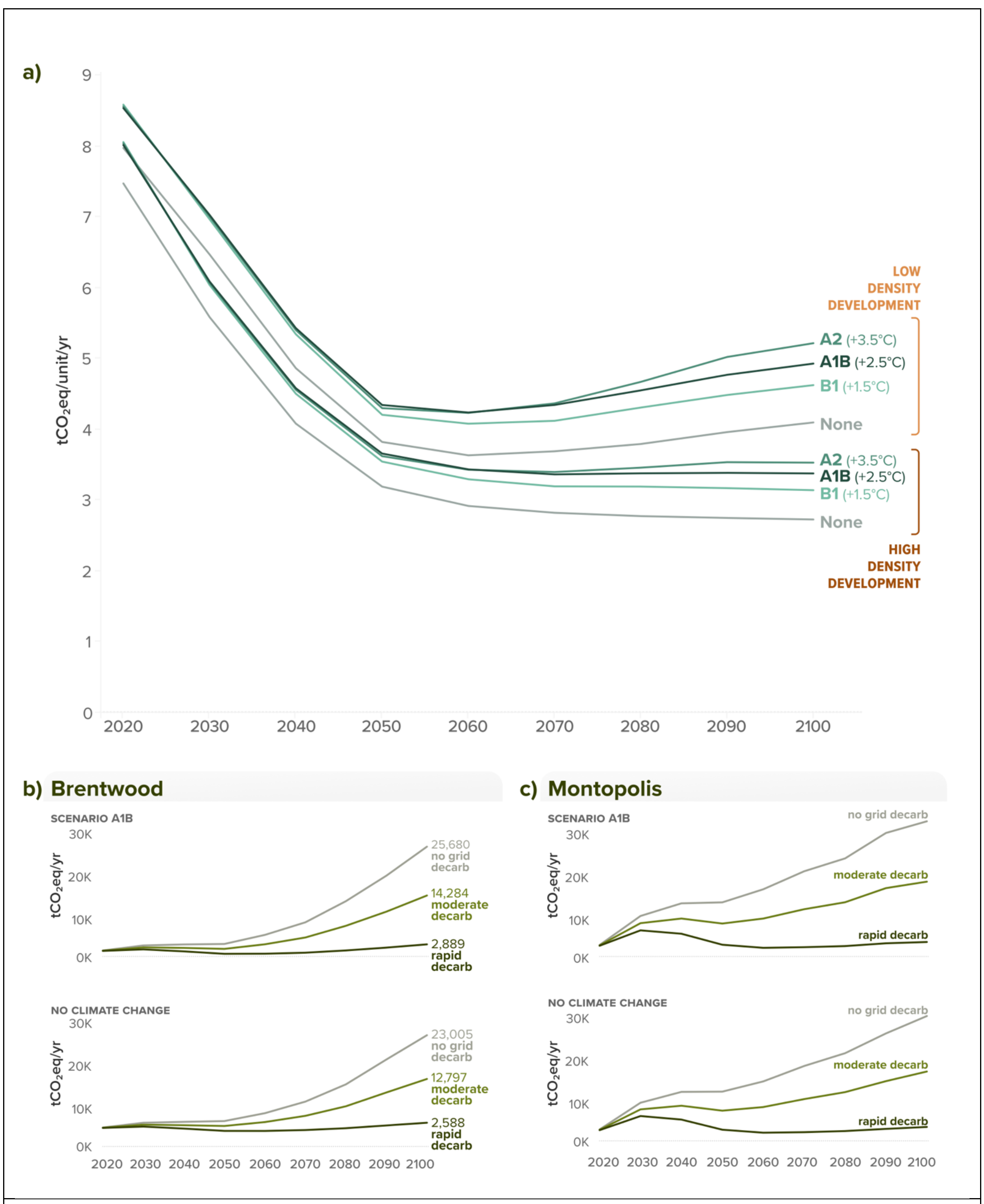

**Figure 8** : a) Impact of climate change and urban redevelopment on relative emissions under moderate grid decarbonization aggregated for all three neighborhoods. Premium for Sprawl for A1B and no climate change in b) Brentwood and c) Montopolis for different grid decarbonization scenarios.

### 3.5 Grid Decarbonization and Urban Development: Premium for sprawl

Figure 8 b) and c) shows the premium for sprawl, i.e., the surplus of emissions, for two neighborhoods for all grid decarbonization scenarios and the A1B and no climate change scenarios (and no technology adoption). For Brentwood (Figure **8** b)) we show the values in 2100 as an example. Note that there is no scenario in which the premium is negative, i.e., high-density development is always more favorable in terms of reducing emissions (for a given population size). Even for rapid grid decarbonization and without considering climate change (Figure 8 b) bottom), in 2100 the low-density development still emits on the order of 2,500 $tCO_2eq$ more annually than the high-density equivalent. For moderate grid decarbonization, the premium for sprawl is about five times higher, ~12,800 tCO2eq, and it is about nine-fold if the grid is not decarbonized (23,000 tCO2eq). These ratios are relatively consistent across the decades and neighborhoods. The A1B climate scenario amplifies the premium for sprawl by about +10%, for example, for the moderate case 14,300 tCO2eq (A1B) compared to 12,800 tCO2eq (no climate change). For B1 and A2 climate scenarios, we find this amplification to be +5% and +15%, respectively.

Comparing (Figure 8 b) and (Figure 8 c), we can also identify the interaction between urban redevelopment and grid decarbonization. The Brentwood neighborhood (Figure 8 b) develops more slowly in the first part of the century, while Montopolis (Figure 8 c) develops more quickly. Consequently, there are relatively fewer new buildings in Brentwood compared to Montopolis, and the difference between the low- and high-density developments is small, reflected in the similar evolution of the premium for sprawl until about 2050. As Montopolis develops more quickly earlier, the premium for sprawl increases in the beginning until about 2040. Eventually, the decarbonization of the grid progresses sufficiently to reduce the premium again. From 2050 onwards both neighborhoods undergo significant redevelopment, highlighting the premium for sprawl for moderate and no decarbonization scenarios, while maintaining comparatively low constant values for the rapidly decarbonizing grid.

### 3.6 . Absolute vs relative emission pathways

For comparison between scenarios, relative values are convenient by normalizing, e.g., for the built area ($tCO_2eq/m^2$), which, in analogy to the energy use intensity index, could also be referred to as emission intensity of a building or neighborhood. Since we are also considering urban development, we put forward that the comparison of emissions per built residence unit ($tCO_2eq$/unit) is more use- and insightful than per built area, since the same built area could potentially serve multiple units. By residence *units,* we understand the subdivisions of a building, i.e., a multi-family house is one building with several housing units.

Relative emission values must be used and interpreted with care as they can obfuscate real pathways. This is shown in **Figure 9** for Brentwood's low-density redevelopment for the moderate grid decarbonization scenarios. All pathways show decreasing emissions until about 2050. After 2050, the annual emissions increase for both ($tCO_2eq$) and ($tCO_2eq$/unit), while they continue to decline for ($tCO_2eq/m^2$). Any indicator can be useful and appropriate depending on the purpose. For example, the absolute emissions indicate the carbon footprint of the neighborhood, while comparisons of interventions for the same urban development scenario can be compared using the built area ($tCO_2eq/m^2$) because the same buildings and residence units are compared to each other.

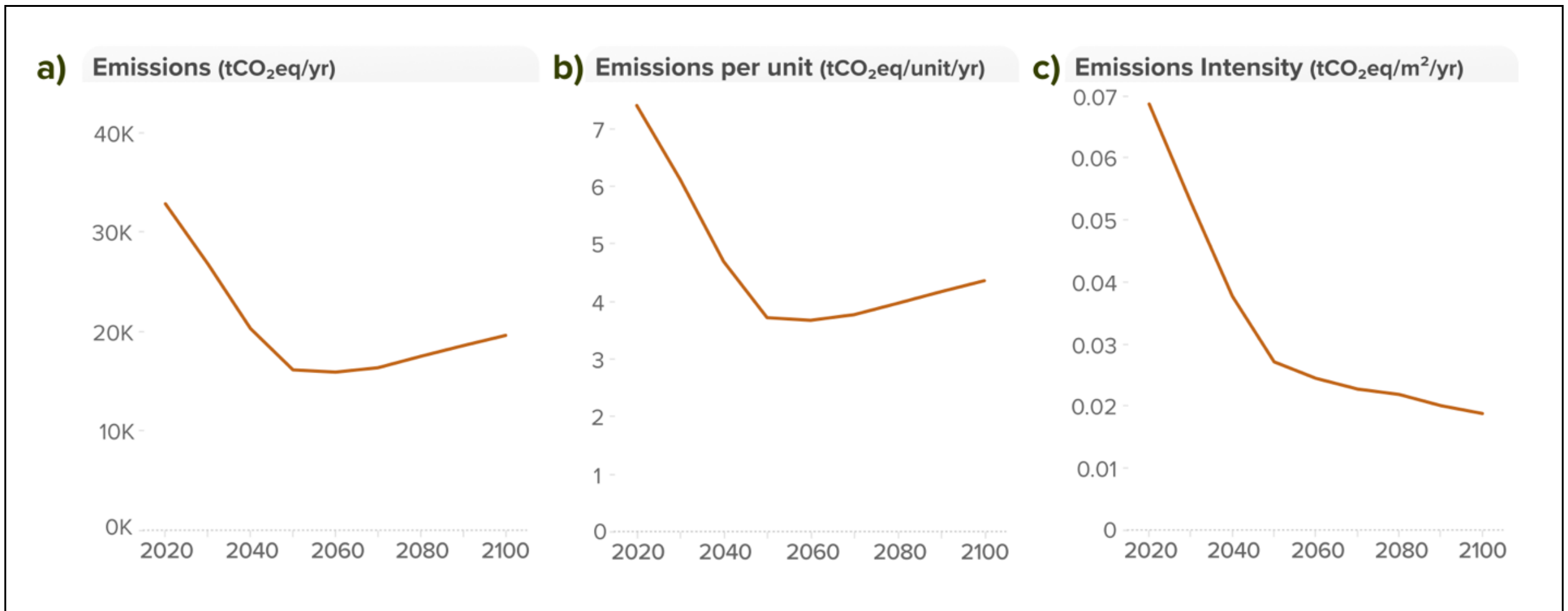

**Figure 9** a) Emission Pathways for Brentwood for climate change scenario A1B in absolute values ($tCO_2eq$), b) per housing unit ($tCO_2eq/unit$), and c) per built area ($tCO_2eq/m^2$).

However, comparisons between different urban redevelopment scenarios should use ($tCO_2eq/unit$) to account for the fact that there exists a cap on the total number of people that a neighborhood can accommodate. This cap is smaller for lower density developments, and so in a generally growing city, additional people must move elsewhere, and their emissions are not captured in absolute pathways comparisons. The *premium for sprawl* indicator resolves the issue of the correct but harder to interpret relative emission pathways using ($tCO_2eq/unit$) and offers a true comparison between different pathways for a neighborhood or municipality.

## 4. DISCUSSION

It is tempting to consider our results of different decarbonization measures independently and rank them based on their ability to reduce emissions. However, this would disregard several long term temporal effects and potential interactions between the measures, which can be discovered in bottom-up exploratory models. Our results clearly show that researchers must include climate change forecasts in energy modeling to ensure that potential shifts in heating and cooling loads are adequately captured. Policy makers must plan with long term scenarios to avoid rebound effects as short term gains can be canceled out by long term developments.

The IMPACT pathways demonstrate that comparatively short-term emission reductions driven by one mechanism, e.g., grid decarbonization, can be effectively overturned in the longer term by a potentially adverse set of events, e.g., unguided urban development. This highlights the necessity to decarbonize the electric grid as fast as possible to avoid adverse effects: if the initially dominant mechanism stalls at some point, e.g., for the moderate grid decarbonization scenario, urban development can eventually counteract all the gains. Furthermore, the IMPACT pathways have shown that in the short-term several combinations of scenarios can potentially produce the same emission reductions, while their impact differs in the longer term. Of course, given the obvious uncertainties associated with making long-term predictions of coupled systems, rigorous monitoring and accounting of emissions is necessary to keep track of model predictions, and develop better models and forecasts as more data is collected.

IMPACT pathways are exploratory and allow us to compare both top-down (policy) and bottom-up (individual) processes to understand the degree and type of interventions necessary to rapidly curb residential emissions. What's more, they also clearly show where the impactful decision-makers are located. For example, in our case study, the individual decision makers, i.e., the technology adopters in the residences, do not emerge sufficiently powerful to have an impact on emissions. Consequently, policy incentivizing or mandating certain technology adoptions, typically at the national level, has equally little impact. Whether this is due to the low number of adopters or low impact of the adopted technology is not clear. By contrast, grid decarbonization and zoning policy have the largest individual impact. The two must be coordinated to effectively deliver an effective net-zero emissions agenda, which will likely heavily involve the municipality level (IEA, 2021).

Densification has long been theorized as central to decarbonization (Teller, 2021), and it has been well established that per capita transportation emissions decline as density rises (Gately, Hutyra, & Wing, 2015). However, there had remained a gap in understanding "the magnitude of the emissions reduction from altering urban form, and the emissions savings from integrated infrastructure and land-use planning" (Intergovernmental Panel on Climate Change, 2015). The IMPACT pathways presented here contribute to uniting "measurement and meaning" in integrated land-use and infrastructure studies (Richter, 2021). We clearly demonstrate that zoning policy and housing can have a substantial impact and must be considered as a viable decarbonization measure. We have introduced the *Premium for Sprawl* to quantify the fact that low density developments accrue emissions outside of their geographical limits due to people who cannot be accommodated in the neighborhoods.

When evaluating our results, one must be aware of the limitations built into the assumptions. It is not clear how our results would be impacted if transportation emissions would be directly included. We can hypothesize that lower density development will be more favorable for transportation emissions but quantifying them requires additional research. An interesting avenue for future work is to integrate the transition to electric vehicles as it would couple transportation emissions directly to the grid. Again, a sprawling development will likely require more frequent EV charging, and therefore result in higher emissions compared to the denser development. Given that our model is at the building unit level, it is relatively straightforward to implement assumptions on EV charging (load and frequency). Similarly, to obtain an even more integrated picture of the impact of the built environment on emissions, embodied emissions should be integrated in future research. This would help compare the impact of efficiency upgrades to new construction.

IMPACT pathways are complimentary to life-cycle assessment (LCA) research on net-zero energy neighborhoods, e.g., (Wiik et al., 2022) in that they provide a temporal exploration of possible scenarios and policy impacts ("what-if" questions), rather than holistic design guidelines, or normative pathways to achieve a certain emission target.

Since we are working with annual average energy demand and grid carbon content, more research is needed to determine the potential of more fine grained temporal resolution for the operation of the electrical grid. It is also not clear what the impact of other potential building efficiency upgrades, e.g.

insulation, would be. However, we can hypothesize that since the impact of high efficiency is modeled by reduced annual energy demand, any equivalently impactful technology will have a similar effect, especially considering the mandate scenario. In that case, the impact is limited by the fraction of buildings redeveloped , which in our model is already relatively large (>6%/yr) compared to the typical 2-3% retrofit rate.

## 5. CONCLUSION

IMPACT pathways generate results consistent with recent studies on the emission reduction potential of buildings (Goldstein et al., 2020) and on the relatively low impact of technology adoption (Creutzig et al., 2021). They integrate decisions and processes at the individual residence level and can be easily further integrated with other policies and impacts of interest, e.g., fuel switching in heating systems, building retrofit (Felkner & Brown, 2020), transportation emissions, adoption of electric vehicles, or embodied carbon in building construction (Berrill & Hertwich, 2021). Further, hourly energy simulation results not presented here, could be used for studies on demand response programs and on grid-interactive buildings (Department of Energy, 2021; Vazquez-Canteli & Nagy, 2019). To unlock the tremendous potential that the built environment offers to address climate change, integrated multi-domain models spanning several spatiotemporal scales can inform decision makers on the effectiveness of policies. IMPACT pathways are key in defining and analyzing policies, as well as in tracking their implementation progress.

## DATA PROCESSING AND AVAILABILITY

The output of each model is organized into sets of spreadsheets in which a given parcel is associated with its own unique `ParcelID`, allowing for linkage of the models. Data are joined and further processed in a set of Tableau workbooks. An interactive dashboard with all the data is available at [REMOVED for anonymity]